\documentclass{aastex}
\usepackage{natbib,graphicx,latexsym}
\usepackage{emulateapj5}

\newcommand{\perone}{\mbox{$^{-1}$}}

\newcommand{\perthree}{\mbox{$^{-3}$}}

\newcommand{\eg}{e.g.}
\newcommand{\ie}{i.e.}

\newcommand{\htwoo}{{\hbox{H$_2$O}}}   
\newcommand{\htwo}{{\hbox{H$_2$}}}     
\newcommand{\meth}{{\hbox{CH$_4$}}}   

\newcommand{\degs}{\mbox{$^{\circ}$}}

\newcommand{\Teff}{\mbox{$T_{\rm eff}$}}

\newcommand{\Rc}{\mbox{${R}$}}
\newcommand{\Ic}{\mbox{${I}$}}
\newcommand{\zp}{\mbox{${z^\prime}$}}

\slugcomment{{\em Astrophysical Journal Letters}, in press}
\shorttitle{Methane Dwarf from the IfA-Deep Survey}
\shortauthors{Liu et al.}

\begin{document}

\title{Discovery of a Methane Dwarf from the IfA-Deep
Survey\altaffilmark{1}} 
\altaffiltext{1}{Based on observations obtained at the Subaru Telescope,
Infrared Telescope Facility, and W. M. Keck Observatory.}


\author{\sc Michael C. Liu\altaffilmark{2}, Richard Wainscoat, Eduardo
L. Mart\'in, Brian Barris, John Tonry} 
\affil{Institute for Astronomy, University of Hawai`i, 2680 Woodlawn
Drive, Honolulu, HI 96822} 
\altaffiltext{2}{Beatrice Watson Parrent Fellow.}
\email{mliu@ifa.hawaii.edu}

\begin{abstract}
We present the discovery of a distant methane dwarf, the first from the
Institute for Astronomy (IfA) Deep Survey.  The object (``IfA~0230-Z1'')
was identified from deep optical \Ic\ and \zp-band imaging, being
conducted as an IfA-wide collaboration using the prime-focus imager
Suprime-Cam on the Subaru 8.2-m Telescope.  IfA~0230-Z1 is extremely red
in the $\Ic{\zp}J$ (0.8--1.2~\micron) bands but relatively blue in
$J-H$; such colors are uniquely characteristic of T~dwarfs.  A near-IR
spectrum taken with the Keck Telescope shows strong \htwoo\ absorption
and a continuum break indicative of \meth, confirming the object has a
very cool atmosphere.  Comparison with nearby T~dwarfs gives a spectral
type of T3--T4 and a distance of $\sim$45~pc.  Simple estimates based on
previous T~dwarf discoveries suggest that the IfA survey will find a
comparable number of T~dwarfs as the 2MASS survey, albeit at a much
larger average distance.  We also discuss the survey's ability to probe
the galactic scale height of ultracool (L and T) dwarfs.
\end{abstract}


\section{Introduction}

In the past few years, observational studies of substellar objects have
been revolutionized, in large part from the advent of wide-field sky
surveys, namely the 2-Micron All-Sky Survey (2MASS;
\citealp{1997ilsn.proc...25S}), the Deep Near-Infrared Survey of the
Southern Sky (DENIS; \citealp{1999A&A...349..236E}), and the Sloan
Digital Sky Survey (SDSS; \citealp{2000AJ....120.1579Y}).  The coolest
of the known substellar objects, the T~dwarfs, are identified by
absorption features from \meth, which becomes the dominant C-bearing
molecule for $\Teff\lesssim1300$~K \citep{1996ApJ...472L..37F}.  The
first T~dwarf, GL~229B, was found as a companion to a nearby M~star
\citep{1995Natur.378..463N}.  Field T~dwarfs have subsequently been
found by SDSS \citep{1999ApJ...522L..61S, 2000ApJ...531L..61T,
2000ApJ...536L..35L, geb01}, 2MASS \citep{1999ApJ...522L..65B,
2000ApJ...531L..57B, 2000AJ....120.1100B, burg01}, and the NTT Deep
Field \citep{1999A&A...349L..41C}.  To date, more than 20 of these
objects are known \citep{burg01b}. These objects are of great interest
for numerous reasons, \eg, they are the lowest mass field objects known
to date, and their physical properties are in many ways more akin to
giant planets than to stars.

The spectral energy distributions (SEDs) of T~dwarfs are distinct from
all other known astronomical bodies, characterized by steeply rising
flux in the optical from $\sim$0.8--1.0~\micron\ due to the
pressure-broadened K~I doublet and a near-IR continuum
($\lesssim2.5~\micron$) showing strong absorption from \htwoo, \meth,
and collisionally induced \htwo\ opacity \citep{1999ApJ...519..802K,
2000ApJ...531..438B, 2000ApJ...533L.155L, burg01, geb01, leg01}.  As a
result, their broad-band colors are unique: their optical colors are
extremely red up to $J$-band (1.25~\micron), while near-IR colors are at
least as blue as early-type stars, with the coolest objects being
uniquely blue in the IR.  Hence, pure infrared surveys such as 2MASS can
identify the coolest T~dwarfs, which have very strong \meth\ absorption,
but are insensitive to the warmer objects, which have IR colors similar
to main sequence stars, giant stars, and asteroids.  In contrast, the
far-red ($i^\prime$ and \zp) bands of the SDSS provide good sensitivity
to T~dwarfs over a wide range of temperture, including ``L/T
transition'' objects \citep[e.g.][]{2000ApJ...536L..35L} where methane
is just beginning to appear in the SEDs.  In addition, the much higher
etendue of current optical imagers compared to near-IR ones indicates
that surveys in the far-red should be very scientifically fruitful.
However, the relatively shallow depth of SDSS means that T~dwarfs can be
found only out to a distance of $\sim$30~pc.

The IfA-Deep Survey is an imaging survey being conducted as
collaboration among many members of the Institute for Astronomy (IfA) at
the University of Hawaii during 2001--2002.  The project uses the
prime-focus imager Suprime-Cam on the Subaru 8.2-m Telescope to map
several blank fields in the far-red, the $RI\zp$ bands
(0.6--1.0~\micron), with a total area of 2.5~sq.\ degrees.  This survey
will serve a variety of scientific programs, including weak lensing;
searches for high-redshift supernovae, extremely red galaxies, and
high-redshift clusters; galactic structure studies; and time-variability
surveys.  Since the survey is on-going, a complete description of its
properties awaits a future paper. The baseline design calls for
10~nights of observing, reaching limiting 5$\sigma$ Vega magnitudes of
$R\approx27.1$, $I\approx 26.5$, and $\zp\approx 25.5$~mags, though the
final outcome will depend on the vicissitudes of telescope performance
and weather.  Here we present the discovery of the survey's first
T~dwarf, ``IfA~0230-Z1'', found from the first clear night of
observations.  We also examine the survey's total T~dwarf yield and its
sensitivity to the vertical scale height of ultracool dwarfs once
completed.

\clearpage
\section{Observations}

\subsection{Subaru/Suprime-Cam: Optical Imaging}

Optical imaging of the RA 02:30~hr field of the IfA-Deep Survey was
obtained on 22 October 2001 UT using the prime focus imager Suprime-Cam
\citep{1998SPIE.3355..363M} on the Subaru 8.2-m Telescope on Mauna Kea,
Hawaii.  The instrument is a mosaic of ten contiguous 2048$\times$4096
MIT/Lincoln Lab phase 2 and 3 CCDs with a total field of view of
34\arcmin$\times$27\arcmin. Conditions were photometric with 0\farcs8
FWHM seeing.  We imaged two adjacent fields with a total area of
0.5~sq.\ degrees.  Dithered \Rc, \Ic, \zp-band observations were
obtained with integrations of 560~s, 645~s, and 960~s per filter per
pointing, respectively.  The \Rc\ and \Ic-band filter are Cousins
filters.  The \zp\ filter has an effective wavelength of 9195~\AA\ and a
FWHM of 1410~\AA, very similar to that used by the SDSS survey
\citep{1996AJ....111.1748F}.  

Images were flattened, defringed, warped onto a common sky coordinate
system, registered, and cleaned of cosmic rays.  A preliminary
photometric calibration of the \Rc\ and \Ic\ data was done using optical
imaging previously obtained from smaller telescopes, which was
calibrated with standards from \citet{1992AJ....104..340L}.  For the
\zp\ data, the preliminary calibration was done by comparing the
\Rc\Ic\zp\ colors of stars in the Suprime-Cam data with the stellar
locus.  The latter was synthesized from the spectral energy
distributions of \citet{1983ApJS...52..121G} and the instrumental
(filter+detector+atmosphere) transmission profiles.  The resulting
magnitudes in Table~1 are Vega-based.

\subsection{IRTF/Spex and Keck/NIRSPEC: Infrared Follow-Up}

We identified an extremely red stellar object at RA(2000) = 02:26:37.6,
DEC(2000) = 00:51:54.7 in the $I$ and \zp-band imaging. We refer to it
as ``IfA~0230-Z1'' hereinafter.  We obtained $J$ and $H$-band photometry
on 06 November 2001 UT using the facility spectrograph Spex
\citep{1998SPIE.3354..468R}.  Spex has a slit-viewing camera, which uses
a 512$\times$512 InSb array from Raytheon-SBRC and has a pixel scale of
0\farcs118~pixel\perone.  Conditions were photometric with 0\farcs85
FWHM seeing.  We obtained a total of 18~min and 10~min of integration at
$J$ and $H$-band respectively.  The Spex filters were purchased as part
of the Mauna Kea Filter Consortium \citep{mkofilters1, mkofilters2}, and
hence will be common to most of the current major infrared
telescopes. We obtained images of the standard star SJ~9105 from
\citet{1998AJ....116.2475P} immediately after observing IfA~0230-Z1.
The resulting magnitudes in Table~1 are Vega-based.
 
We obtained an $H$-band spectrum of IfA~0230-Z1 on 10 November 2001 UT
with the Keck Telescope and the facility spectrograph NIRSPEC
\citep{1998SPIE.3354..566M}.  NIRSPEC uses a 1024$\times$1024 InSb
ALADDIN detector from Raytheon-SBRC. A total of 30~min of integration
was obtained in low resolution mode using the NIRSPEC-5 blocking filter
and the 0\farcs76 slit. Conditions were very non-photometric due to high
thick cirrus. The object was dithered on the slit between exposures.
The slit PA was set to 73\fdg6 east of north, so that the bright object
5\farcs65 away (``object~A'') was also on the slit for all the
exposures.  This provided a well-detected reference for registering the
frames, and also a check on the resulting spectrophotometry (see below).
A nearby A0V star was observed immediately afterward to calibrate the
telluric and instrumental throughput.

The spectra were reduced using custom IDL scripts.  The raw images on
the NIRSPEC detector are curved in both the spectral and spatial
directions.  After subtracting a dark frame and dividing by a flat
field, the individual images were cleaned of outlier pixels and
rectified using traces of the arc lamp lines and the object spectra.
Pairs of images taken at successive nods were subtracted to remove the
sky emission.  Images were then registered, shifted, and stacked to form
a final 2-d mosaic of the spectrum.  Extractions of 1-d spectra from the
mosaic were done in a manner to produce reliable errors based on photon
counting (Poisson) statistics.  Details will be presented in a future
paper.  We divided the extracted spectra by the spectra of the A0V
calibrator star and then multiplied by a 9720~K blackbody to restore the
true shape of the continuum.  Hydrogen absorption features in the
calibrator were removed by linear interpolation.  Wavelength calibration
was done with spectra of argon and neon lamps.

The spectral resolution ($\lambda/\Delta{\lambda}$) of the original
extracted spectra was $R=1640$ (9.7~\AA); many of the telluric OH
emission lines are well-separated from each other.  The resulting S/N
was only $\approx1\!-\!3$~pixel\perone, so we smoothed the spectra with
a 32~pixel FWHM Gaussian, with proper weighting accounting for the
measurement errors, and rebinned the data to 2 pixels per spectral
resolution element.  The final spectra have a resolution of $R=180$ and
are plotted in Figure~\ref{fig-spectra}.


\section{Analysis}

\subsection{Spectral Classification}

The \Ic\zp$J$-band colors of IfA~0230-Z1 are extremely red, while the
$J-H$ colors are relatively blue.  Such colors are uniquely
characteristic of T~dwarfs.  IfA~0230-Z1 has $\zp-J=2.74\pm0.15$~mag
(Vega), comparable to the reddest known T~dwarfs.  Its $J-H$ color of
0.34$\pm$0.05 indicate a spectral type of T3--T4 \citep{leg01}.

\begin{figure}[t]
\hskip -1.6in
\hbox{
\raise 0.25in
\vbox{
\centerline{\includegraphics[width=3in]{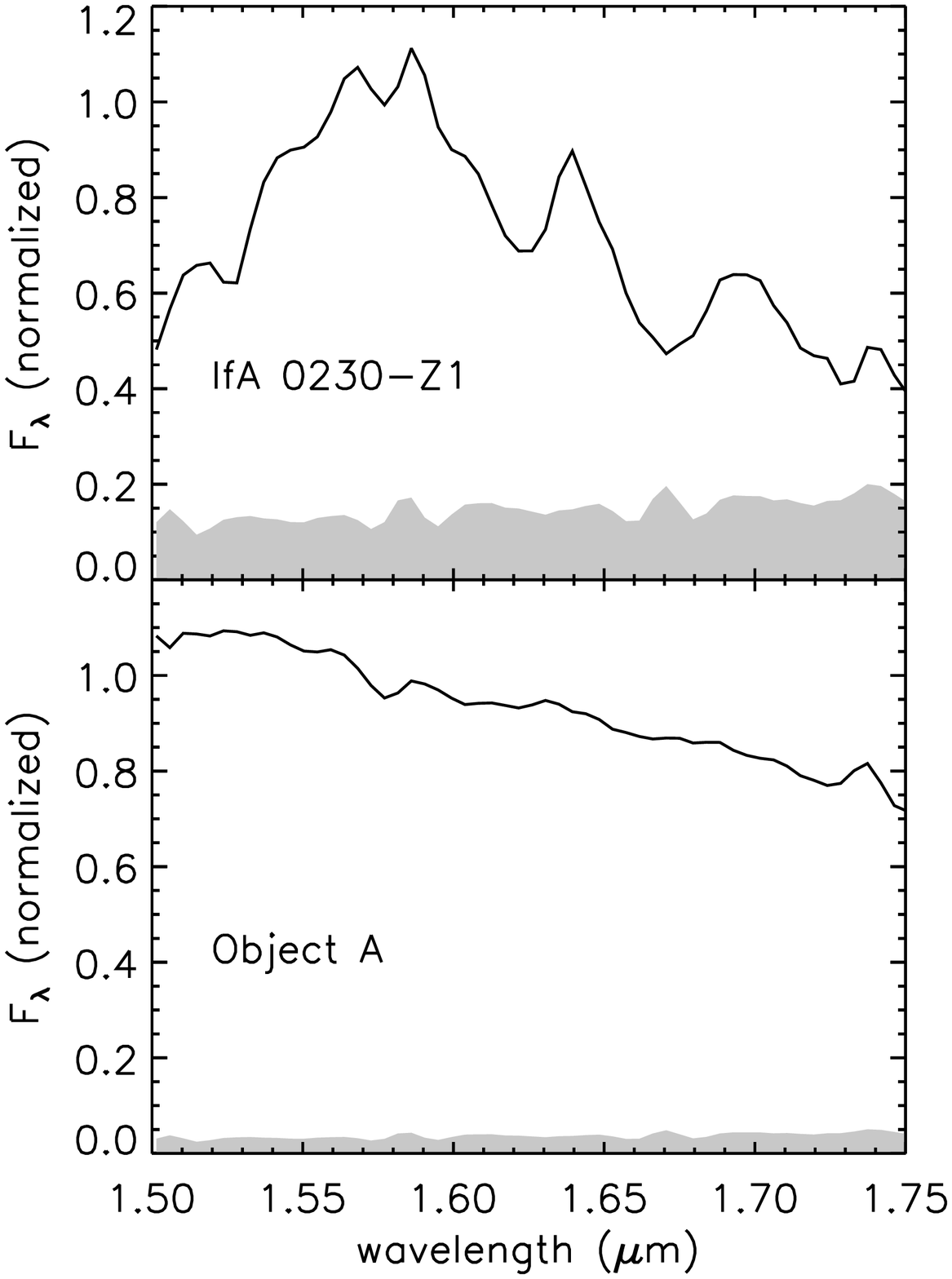}}}
\hskip -4.5in  
\centerline{\includegraphics[width=3.3in]{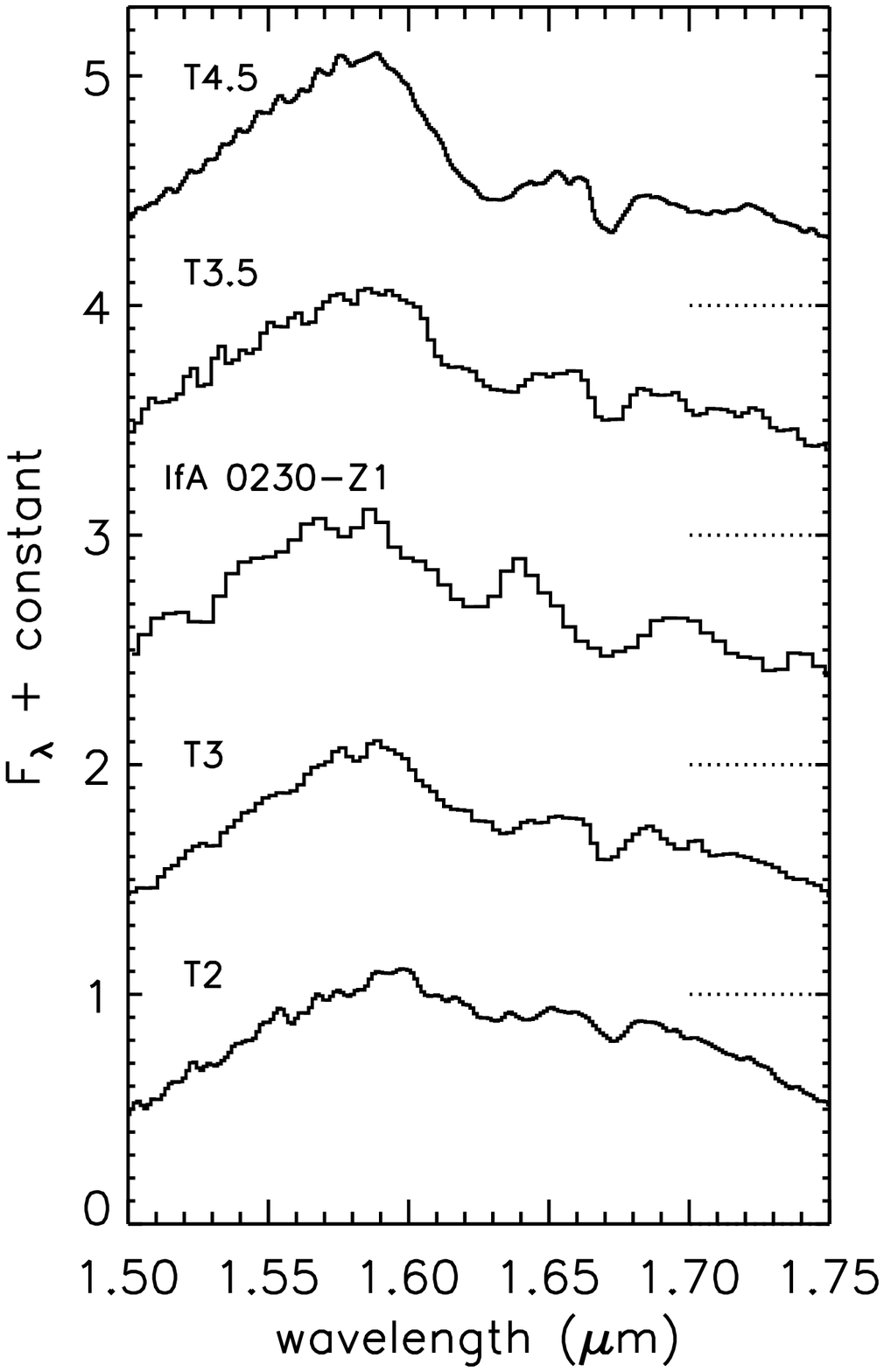}}}
\caption{\normalsize {\bf Left:} Keck/NIRSPEC spectra of IfA~0230-Z1 and
an adjacent bright galaxy observed simultaneously on the same slit.  The
spectra have been smoothed to $R=180$ and rebinned to 2 pixels per
spectral resolution element.  The formal 1$\sigma$ errors are shown as
shaded regions at the bottom of each plot.  IfA~0230-Z1 shows a peak
near 1.58~\micron\ and depressed flux on the blue and the red sides,
indicative of \htwoo\ and \meth\ absorption, respectively.  In contrast,
object~A shows a featureless continuum.
{\bf Right:} IfA~0230-Z1 compared with spectra of local T~dwarfs from
\citet{geb01}: SDSS~1254 (T2), SDSS~1021 (T3), SDSS~1750 (T3.5),
2MASS~0559 (T4.5). Each spectrum has been normalized to an average flux
of 1.0 in the 1.55--1.60~\micron\ region and then offset by an integer
value.  The dotted lines indicate the zero flux level for each spectrum.
\label{fig-spectra}}
\end{figure}

Figure~\ref{fig-spectra} shows the Keck/NIRSPEC spectra of IfA~0230-Z1
and object~A, which were observed simultaneously and reduced in an
identical fashion.  The spectrum of object~A shows a featureless
continuum $f_\lambda\sim\lambda^{-0.7}$, consistent with the near-IR
continuum of a low redshift galaxy \citep{2001MNRAS.326..745M}.  On the
other hand, IfA~0230-Z1 shows a peak in its continuum around
1.58~\micron. The continuum is depressed in the blue and the red around
the peak, indicating the presence of \htwoo\ and \meth\ absorption,
respectively.  Since object A's spectrum does not show any such
features, we conclude that these features are real and not due to, \eg,
systematic errors in the spectrophotometry due to telluric water
vapor.\footnote{We also observed IfA~0230-Z1 using Keck/NIRSPEC with the
same instrumental configuration on the previous night for a total of
20~min.  Conditions were more cloudy, and the resulting spectrum had
much lower S/N. However, the overall shape of the object's continuum was
consistent with the higher S/N data in Figure~\ref{fig-spectra}, with a
factor of $\sim$2 drop from the peak to the reddest wavelengths.  Given
its very low S/N, we chose not to use this data.}

Figure~\ref{fig-spectra} provides a comparison of IfA~0230-Z1 with local
T~dwarfs found in the SDSS and classified by \citet{geb01}.  Based on
this, we estimate a spectral type of T3--T4 for IfA~0230-Z1, which
agrees with the typing from the broad-band colors alone.  Since the
shape of the $H$-band continuum is changing rapidly with spectral type,
the uncertainty in the spectral typing is no more than 1~spectral
subclass.

\subsection{Distance}

Only three T~dwarfs have known distances, all of them companions to main
sequence stars.  Two of them, GL~229B and GL~570D, are much cooler
objects than IfA~0230-Z1, with stronger \meth\ absorption
\citep{1995Sci...270.1478O, 2000ApJ...531L..57B}.  The third, GL~86B,
appears to have modest methane absorption like IfA~0230-Z1, but its
magnitudes and colors are poorly known \citep{els01}.  Also, at fixed
effective temperature/spectral type, younger (less massive) brown dwarfs
will be more luminuous than older (more massive) ones.  Hence, even if
there were T3~dwarfs with known distances, their ages and hence their
luminosities might be different than IfA~0230-Z1.

We estimate the distance as follows.  For ages older than $\sim$0.1~Gyr,
the radius of a brown dwarf is largely independent of its mass, to
within about 30\% \citep{bur01}.  Hence, $L\sim\Teff^4$.  The bolometric
correction at $J$-band is observed to be nearly constant for late-L and
T~dwarfs \citep{leg01} so the absolute $J$-band magnitude $M_J$ will
scale directly with \Teff$^4$.  The difference in effective temperature
between the late-L dwarfs and the late-T dwarf GL~229B is estimated to
be quite small, perhaps only 1300 to 1000~K \citep{2000AJ....120..447K,
burg01}.  Adopting $\Teff\approx1150$~K for IfA~0230-Z1 means its
$J$-band absolute magnitude will be $\approx$0.6 mag brighter than for
GL~229B, which has $M_J = 15.51\pm0.09$~mag \citep{1999ApJ...517L.139L}.
(Note that this is a differential comparison and does not depend on the
absolute \Teff\ scale adopted.)  Hence, we estimate $M_J\approx14.9$~mag
for IfA~0230-Z1. This compares favorably to the two L8~dwarfs with known
distances, which have an average $M_J\approx14.85$~mag
\citep{2000AJ....120..447K}.  The resulting distance estimate for
IfA~0230-Z1 is 45~pc, with an error of $\sim$20\% ($\pm$9~pc) based on
the uncertainties in the radius and absolute $J$-band magnitude.

\subsection{Survey Predictions: Number Counts and Scale Height
of Ultracool Dwarfs}

The 0230 field of the IfA-Deep survey is contained with the SDSS Early
Data Release \citep{sdss-edr}, but IfA~0230-Z1 is $\sim$0.6~mags too
faint in \zp\ to be detected by SDSS.  With the detection of only a
single object, any discussion of the T~dwarf number counts at magnitudes
fainter than SDSS is unwarranted.  However, we can do a simple
comparison.  The first five T~dwarfs from SDSS
\citep{1999ApJ...522L..61S, 2000ApJ...531L..61T, 2000ApJ...536L..35L}
were found in an area of 355~square degrees, with an effective limiting
magnitude of $\zp\approx19.5$~mags (Vega).  This means a surface density
of about 1~per 70 square degrees. Our \zp-band imaging of the 0230 field
used in this paper reaches $\sim$4 mags deeper. Assuming a uniform
volume density of T~dwarfs, the expected surface density would be 1~per
0.3~square degrees, consistent with our discovery of a single object.
The final IfA survey will go a factor of 3 deeper in flux and cover 5
times more area, suggesting a total yield of $\sim$40 T~dwarfs. In
comparison, 13 T~dwarfs have been found from the $\sim$40\% of the 2MASS
data searched to date \citep{burg01b}.  This suggests the IfA survey
will find a comparable number of T~dwarfs as the entire 2MASS survey,
albeit at a much larger average distance.  However, these numbers should
be taken with caution since they are based on the small sample of
objects found to date by SDSS.

The completed IfA-Deep survey will be sensitive to T~dwarfs out to
$\sim$300~pc and L~dwarfs out to $\sim$2~kpc.  By probing several lines
of sight, the survey should provide the first insights into the vertical
scale height of ultracool dwarfs.  To examine this aspect quantiatively,
we consider a simple model of an exponential disk based on
\citet{1992ApJS...83..111W}, with a radial scale length of 3.5~kpc and
different vertical scale heights.  For the L~dwarfs, we adopt a local
volume density of 0.01~pc\perthree\ and assume an equal number of
objects per spectral subclass (L0 to L8), consistent with the analyses
of \citet{1999ApJ...521..613R} and \citet{chab01}.  We adopt a local
volume density for T~dwarfs of 0.006~pc\perthree, using the
\citet{burg01b} results for T5 to T8 dwarfs and then doubling it to
account for early T~dwarfs excluded from their 2MASS-selected sample.
For the T~dwarfs, we assume half the population is in early T dwarfs,
with the rest equally divided into T6, T7, and T8 dwarfs.  These inputs
are very approximate, but consistent with the limited current
observations.  Finally, we compute \zp-band absolute magnitudes using
data from \citet{2000AJ....120..447K}, \citet{2000ApJ...531L..57B},
\citet{els01}, and \citet{leg01}.

The differences in the observed counts of L and T~dwarfs between our low
and high galactic latitude fields will be very sensitive to the objects'
vertical scale height.  The 0230 field lies at very high galactic
latitude ($b=-53\degs$), while other fields in the IfA-Deep Survey cover
lower latitudes, down to the 0749 field at $b=18\degs$.  For a canonical
scale height of 325~pc, we predict the counts of L~dwarfs in our high
and low latitude fields will differ by a factor of 3.  In contrast, if
the scale height for ultracool objects is 100~pc, akin to the
Population~I constitutents of the Galactic disk
\citep{2000asqu.book.....C}, the difference between high and low
latitude counts will be much larger, about a factor of 9.  For the
T~dwarfs, the dependence on scale height has a different behavior, since
our survey is only sensitive to much closer objects.  For a 325~pc scale
height, the low latitude field is predicted to have about 1.5 times as
many T~dwarfs as the high latitude field. However, in the case of a
100~pc scale height, there should be 3 times as many T~dwarfs in the low
latitude field compared to the high latitude field.


\section{Conclusions}

We have found a methane dwarf from the first night of observations for
the IfA-Deep Survey.  The object has very red optical colors and
relatively blue near-IR colors. Its $H$-band spectrum shows strong
\htwoo\ and modest \meth\ absorption. The inferred spectral type is
T3--T4.  The majority of T~dwarfs known to date have very deep \meth\
absorption, so IfA~0230-Z1 is an interesting object in that it lies in a
transition region of spectral type (and hence \Teff), where methane is
beginning to dominate the SED.  The colors of early T-type objects, like
IfA~0230-Z1, make them difficult to identify in pure near-IR surveys
such as 2MASS but they are easily distinguished in our deep
optical/far-red survey.  Such objects are useful to understand the
physics of ultracool atmospheres. Slightly warmer objects, the late
L-dwarfs, have extremely red near-IR colors, which have been interpreted
as being due to the role of dust.  However, the very blue near-IR colors
of T~dwarfs are consistent with ``clear'', \ie, dust-free, atmospheres
\citep[e.g.][]{2001udns.conf....9T}.

With its high sensitivity in the far-red bands, the IfA-Deep survey will
be an excellent means to search for brown dwarfs.  Preliminary estimates
based on the T~dwarf discoveries by SDSS suggests the IfA survey will
find a comparable number of T~dwarfs as the 2MASS survey, albeit at a
much larger average distance.  The final survey will go $\gtrsim$100
times fainter than SDSS, raising the possibility of finding isolated
brown dwarfs much cooler than those found to date.  Furthermore, the
survey will be sensitive to T~dwarfs out to $\sim$300~pc and L~dwarfs
out to $\sim2$~kpc along several lines of sight, at galactic latitudes
ranging from $|b|\approx20-60\degs$.  This raises the possibility of
providing the first insights into the vertical scale height of ultracool
dwarfs, thereby telling us about their origin in a galactic context.

\acknowledgments

The authors wish to extend special thanks to those of Hawaiian ancestry
on whose sacred mountain we are privileged to be guests.  Without their
generous hospitality, none of the observations presented herein would
have been possible.  It is a pleasure to acknowlege our collaborators on
the IfA-Deep Survey: Herve Aussel, Ken Chambers, Pat Henry, Nick Kaiser,
Eugene Magnier, Dave Sanders, and Alan Stockton.  We are grateful for
the support from the staffs of the Subaru Telescope, IRTF, and Keck
Observatory, including Yutaka Komiyama, Ryusuke Ogasawara, Paul Hirsh,
John Rayner, David Sprayberry, and Gary Puniwai.  We thank Pete Challis,
Alan Tokunaga and Jonathan Leong for help with the Subaru and Keck
observations; Sungsoo Kim, Lisa Prato, and Ian McLean for providing the
REDSPEC reduction software and documentation; and Mamori Doi for SDSS
filter data. The Subaru Telescope is operated by the National
Astronomical Observatory of Japan.  The IRTF is operated by the
University of Hawaii under a contract from National Aeronautics and
Space Administration (NASA).  W.M. Keck Observatory was made possible by
the generous financial support of the W.M. Keck Foundation and is
operated by Caltech, the University of California, and NASA.  M. Liu is
grateful for research support from the Beatrice Watson Parrent
Fellowship at the University of Hawai`i.

\clearpage


\begin{figure}
\vskip -3in
\hskip -0.25in
\centerline{\includegraphics[width=5.5in,angle=90]{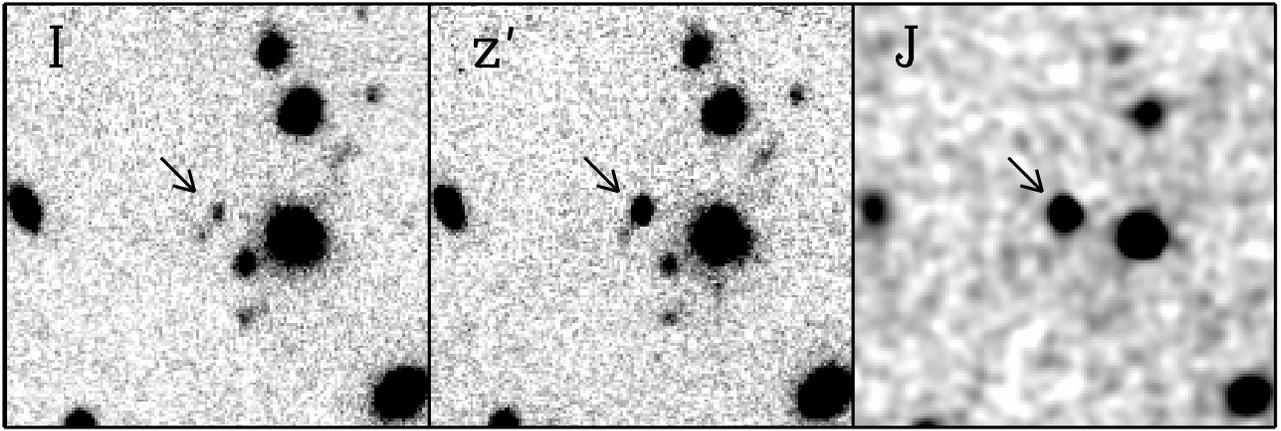}}
\caption{\normalsize Optical and near-IR imaging for IfA~0230-Z1,
indicated by the arrow.  Each image is 30\arcsec\ on a side with North
up and East left. The optical images are slightly trailed due to a
telescope guiding problem. The \Ic\ and \zp-band images are from
Subaru/Suprime-Cam.  The $J$-band image is from IRTF/Spex and has been
smoothed using a gaussian with FWHM equal to the seeing to enhance the
faint objects.  The bright object at PA 286\degs\ and a separation of
5\farcs7 is ``object A,'' used for the Keck/NIRSPEC spectroscopy (see
Figure~\ref{fig-spectra}).
\label{fig-multipanel}}
\end{figure}

\clearpage 

\begin{deluxetable}{lc}
\tabletypesize{\normalsize}
\tablecaption{Photometry of IfA~0230-Z1 \label{table-phot}}
\tablewidth{0pt}
\tablehead{
\colhead{Filter ($\lambda_c$)} & \colhead{Mag (Vega)}}

\startdata
\Rc\ (0.66~\micron)     & $>$25.5 \\
\Ic\ (0.81~\micron)     & 23.61 $\pm$ 0.10 \\
\zp\ (0.92~\micron)     & 20.91 $\pm$ 0.15 \\
$J_{MK}$ (1.24~\micron) & 18.17 $\pm$ 0.03 \\
$H_{MK}$ (1.65~\micron) & 17.83 $\pm$ 0.04 \\
\enddata



\end{deluxetable}

\end{document}